\documentclass[aps,prl,twocolumn,superscriptaddress]{revtex4}
\usepackage{graphicx}
\usepackage[hypertex]{hyperref}

\def\beq{\begin{equation}}
\def\eeq{\end{equation}}
\def\bea{\begin{eqnarray}}
\def\eea{\end{eqnarray}}

\def\kms{~{\rm km/s}}

\def\msun{M_{\odot}}
\usepackage{color}

\begin{document}

\title{Hierarchy in the Phase Space and Dark Matter Astronomy}

\author{Niayesh Afshordi}\email{nafshordi@perimeterinstitute.ca}
\affiliation{Perimeter Institute
for Theoretical Physics, 31 Caroline St. N., Waterloo, ON, N2L 2Y5,Canada}
\affiliation{Department of Physics and Astronomy, University of Waterloo, 200 University Avenue West, Waterloo, ON, N2L 3G1, Canada }
\author{Roya Mohayaee}\affiliation{Institut
d'Astrophysique de Paris, CNRS, UPMC, 98 bis boulevard Arago, Paris, France}
\author{Edmund
Bertschinger}\affiliation{Department of Physics
and Kavli Institute for Astrophysics and Space
Research, MIT\\ Room 37-602A, 77 Massachusetts Ave., Cambridge, MA 02139, USA}

\date{\today}
\preprint{astro-ph/yymmnnn}

\begin{abstract}
We develop a theoretical framework for describing the
hierarchical structure of the phase space of cold dark matter
haloes, due to gravitationally bound substructures.
Because it includes the full hierarchy
of the cold dark matter initial conditions and is hence complementary to
 the halo model, the stable
clustering hypothesis is applied for the first time here to the
 small-scale phase space structure.
As an application,
we show that the particle dark matter annihilation signal
could be up to two orders of magnitude larger than that of the smooth
 halo within
the Galactic virial radius. The local boost is inversely proportional to the smooth
 halo density, and
thus is ${\cal O} (1)$ within the solar radius, which could translate
 into interesting signatures for dark matter direct detection experiments:
The temporal correlation of dark matter detection can change by a factor
 of $2$ in
the span of 10 years, while there will be significant correlations in
 the velocity
space of dark matter particles. This can introduce ${\cal O} (1)$
 uncertainty in the
direction of local dark matter wind, which was believed to be a
 benchmark of directional
dark matter searches or the annual modulation signal.

\end{abstract}

\maketitle

Among the favorite dark matter candidates are the
weakly interacting massive particles (WIMP), yet to be detected in
particle accelerators.
Different experiments look for a signature of WIMPs directly
as they
pass through Earth and recoil off the atomic nuclei in laboratory
detectors, or indirectly through the byproducts
of their self-annihilation into standard model particles (such as photons,
electron/positron
pairs, or neutrinos). The cross-section for the (s-wave)
self-annihilation is also
fixed by the relic abundance: $\langle \sigma_{\rm ann} v\rangle
\sim 10^{-31} - 10^{-26} {\rm cm^3/s} $.
Therefore, one
would expect an astrophysical luminosity of:
\beq
\hspace{-0.2cm}\frac{dL_i}{dE_i} = \frac{\langle \sigma_{\rm ann} v\rangle E_i}
{2m^2_{\chi}}\frac{dB_i}{dE_i} \Phi\,\;;
\qquad\Phi \equiv \int  \rho^2({\bf x}) d^3{\bf x}\;
\label{dlde}
\eeq
where
$m_{\chi}$ is the WIMP mass, $E_i$ is the
energy of
the byproducts, $dB_i/dE_i$ is the differential branching ratio
into $i$ (=$\gamma, e^+e^-$ or $\nu\bar{\nu}$) particles and
$\Phi$ is fixed by the spatial density distribution of
dark matter within the emitting region.

All the factors in (\ref{dlde}), except for $\Phi$, are fixed by the
particle physics model (e.g. \cite{Gondolo:2004sc}).
Unfortunately the gravitational potential of
dark matter haloes says little
about the contribution of small scale structure to $\Phi$. The
cold dark matter (CDM) primordial power spectrum
predicts a large range of mass scales, from $10^{12}-10^{14} \msun$ down to
$10^{-12}-10^{-4} \msun$ \cite{Profumo:2006bv}, far below the resolution
 limit of the present-day simulations
at $z=0$, which is at best $10^4 \msun$. In
fact, simulations already see a nearly constant
contribution to $\Phi$ per decade in sub-structure mass
\cite{Diemand:2006ik,2008Natur.456...73S}, suggesting a
significant contribution from unresolved structures.

Extrapolation of the simulated
properties of the sub-haloes to below the resolution limit,
as well as assumptions about the spatial/mass
dependence of the boost due to sub-structure
(e.g. \cite{Strigari:2006rd,Pieri:2007ir,Kuhlen:2008aw})
could lead to significant
over/underestimations (especially for a non-scale-invariant
linear power spectrum such as CDM). Similarly, the
assumption of Maxwell-Boltzmann velocity
distribution in direct detection experiments of dark matter particles
could miss important phenomenological
signatures of clustering in the phase space.

In this letter, we use the {\it stable clustering hypothesis} in order
to predict the clustering of dark matter particles in the phase space,
and consider the implications for indirect and direct dark matter
searches. The stable clustering hypothesis was first introduced by
Davis \& Peebles  \cite{1977ApJS...34..425D} as an analytic technique to
study the galaxy correlation function in the deeply non-linear regime,
and was subsequently applied into fitting formulae for non-linear
correlation functions (e.g.
\cite{Hamilton:1991es,Mo:1995db,Peacock:1996ci}). The hypothesis assumes
that the number of neighbors within a fixed physical separation becomes
a constant (or pairwise velocity vanishes) on small scales, when the
non-linear structure formation is completed. However, it became clear
that this cannot be a good approximation on the scale of virial radius
of CDM haloes ($\sim 1~ {\rm Mpc}$ today) as halo mergers and the
subsequent tidal disruption can dissolve the old structures into newly
formed haloes (e.g. \cite{2000ApJ...538L.107M,Smith:2002dz}). Nevertheless,
the alternative analytic framework, known as the {\it halo model}, which
has been extensively used in the literature over the past decade, misses
the small scale structure of the haloes (i.e. sub-haloes,
sub-sub-haloes, etc.) that do naturally form in hierarchical structure
formation, and are ubiquitous in high resolution N-body simulations
(e.g. \cite{2008Natur.456...73S}).

In contrast, the merging and tidal activity should cease at scales much
smaller than
the virial radius of CDM haloes, suggesting that the stable clustering
regime might be
achieved on small enough scales. In other words, while sub-haloes could
lose a large fraction
of their mass due to tidal heating/stripping, a small fraction could
remain gravitationally
bound (e.g. \cite{Goerdt:2006hp}). As the mean/virial density of haloes
drops as 1/time$^{2}$, the  gravitationally bound remnant will
eventually become
resilient to tidal disturbances. The stable clustering hypothesis
can be trivially
extended to the phase space where similar to real
space it would predict that the number of
particles within the physical velocity $\Delta {\bf v}$ and
physical distance $\Delta {\bf r}$ of a given particle does not
change with time for small enough $\Delta {\bf v}$ and $\Delta {\bf
r}$.

\textcolor{black}{
In order to develop the stable clustering formalism in phase space,
we start with the collisionless Boltzmann equation at
the phase space coordinates,
${\bf r} +\Delta{\bf r},{\bf v}+\Delta{\bf v}$, {\it i.e.}
\bea
&&\!\!\!\!\!\!\!\!\!\!{df\over dt}({\bf r} +\Delta{\bf r},{\bf v}+\Delta{\bf v}, t)=\nonumber\\
&&
\!\!\!\!\!\!\!\!\!\!{\partial f\over \partial t}+
{\partial f\over \partial {\bf r}}\cdot({\bf v +\Delta v})-
{\partial f\over \partial {\bf v}}\cdot({\bf{\bigtriangledown}}\phi+
{\bf{\bigtriangledown}}{\bf{\bigtriangledown}}\phi\cdot \Delta{\bf r})=0.
\label{boltzmann}
\eea
The above equation in terms of the new function
\beq
{\tilde f}_i(\Delta{\bf r},\Delta{\bf v})\equiv
f({\bf r}_i +\Delta{\bf r},{\bf v}_i+\Delta{\bf v})
\eeq
for particle $i$, can be re-written as
\beq
{d f\over dt}=
\left. {\partial {\tilde f}_i\over \partial t}\right|_{\Delta {\bf r},\Delta {\bf v}}+
{\partial {\tilde f}_i\over \partial \Delta {\bf r}}\cdot\Delta {\bf v}
-{\partial {\tilde f}_i\over \partial \Delta {\bf v}}\cdot(
{\bf{\bigtriangledown}}{\bf{\bigtriangledown}}\phi\cdot \Delta{\bf r})=0
\label{bolt}
\eeq
where
\beq
\left. {\partial {\tilde f}_i\over \partial t}\right|_{\Delta {\bf r},\Delta {\bf v}} =
{\partial {\tilde f}_i\over \partial t}+{\bf v}\cdot {\partial{\tilde f}_i\over \partial\Delta{\bf r}}
-{\bf{\bigtriangledown}}\phi \cdot {\partial{\tilde f}_i\over \partial\Delta{\bf v}}\;.
\eeq
The stable clustering hypothesis assumes that the
above expression when averaged over the particles vanishes for small $\Delta r$ and $\Delta v$.
If we assume that $\langle \tilde f_i {\bf{\bigtriangledown}}
{\bf{\bigtriangledown}}\phi\rangle_p \approx \langle \tilde f_i \rangle_p\langle
{\bf{\bigtriangledown}}{\bf{\bigtriangledown}}\phi\rangle_p$
then
a solution to
 (\ref{bolt}), averaged over particles is
\beq
\langle\tilde f\rangle_p \equiv \frac{1}{N} \sum_i \tilde{f}_i = F\left[\Delta{\bf v}^2+
\Delta x_j\Delta x_k\langle\partial_j\partial_k\phi\rangle_p\right]\;.
\eeq
which is the most general time-independent solution with an isotropic velocity distribution, where $F$ is an arbitrary function, and $N$ is the number of particles in the phase-space volume of interest.
We later show that averaging over particles and volume-averaging
differ only by a constant.
Next, we use the approximation that
the potential is spherically symmetric and hence
the above solution, using Poisson equation, can be re-written as
\beq
\langle\tilde f\rangle_p=\mu \xi_s=F\left[(\Delta{\bf v})^2+
100 H(\xi_s)^2(\Delta {\bf r})^2\right]\;,
\label{fxis}
\eeq
where $\xi_s$ is the phase-space density at the formation time of the sub-structure, and we have used
the fact that the
post-collapse halo
density is roughly $\sim 200$ times the critical density
at the formation time \cite{Gunn:1972sv}. $\mu \sim 1-10\%$ is the mean fraction of bound particles that
can survive the tidal disruption period.
}

\textcolor{black}{
In order to find the function $F$, we use
the spherical collapse results
\beq
\xi_s \sim
\frac{10H(\xi_s)}{G^2 M(\xi_s)}\;,
\label{xis}
\eeq
and also that the
radius and velocity dispersion of haloes are related by (e.g.
\cite{Afshordi:2001ze}):
$
\sigma_{\rm vir} \sim 10 H r_{\rm vir}\;,
$
where $H$ is the Hubble
constant at the time of halo's collapse.
}
\textcolor{black}{
Hence (\ref{xis}) is roughly the
phase space density of haloes that collapse at Hubble constant
$H(\xi_s)$ and mass $M(\xi_s)$.
The phase space volume of the collapsed halo, i.e. the volume of
the constant-$\xi_s$ ellipsoid in (\ref{fxis}) is $M/\xi_s$ and
using (\ref{xis}) we have:
\beq
\left[\frac{\pi F^{-1}(\mu \xi_s)}{10 H(\xi_s)}\right]^3= {\left[ G
M(\xi_s)\right]^2\over 10 H(\xi_s)}\;.
\eeq
Furthermore, the mass scale that collapses at a given cosmological
epoch is characterized by:
\beq \left[H(\xi_s)\over H_0\right]^{-2/3}
\sigma\left[M(\xi_s)\right] \sim \delta_c \simeq 1.7\;,
\label{deltac}
\eeq
where $\delta_c$ is the linear density threshold
for the spherical collapse, while $\sigma[M]$ is the r.m.s. top-hat linear
overdensity at the mass scale $M$, and $H_0$ is the present day
Hubble constant.
We remark that the effect of dark energy will be a
constant factor that could be absorbed in the definition of $\mu$,
since most sub-haloes have formed long before the  era of dark energy   dominance.
}

\textcolor{black}
{
Using the above results, the phase space correlation function takes the form:
\bea
&&
\!\!\!\langle f({\bf r}_1,{\bf v}_1) f({\bf
r}_2,{\bf v}_2) \rangle \nonumber\\&&\simeq
{1\over V_6}\int_{V_6} d^3r ~d^3v\,\, f({\bf r},{\bf v})\, f({\bf r}+\Delta{\bf r},{\bf v} +\Delta {\bf v})\nonumber\\
&& =
{1\over V_6}\sum_i f({\bf r}_i+\Delta {\bf r},{\bf v}_i+\Delta{\bf v})={N\over V_6}
\langle \tilde f \rangle_p\\
&&\simeq
\langle f({\bf r}_1,{\bf v}_1)
\rangle \langle f({\bf
r}_2,{\bf v}_2) \rangle+  \mu
\langle f(\bar{{\bf r}},\bar{{\bf v}}) \rangle
\xi_{s}(\Delta{\bf r},\Delta{\bf v}),
\nonumber\\
\label{stable}
\eea
where we used the assumption of ergodicity to replace the ensemble average $\langle\rangle$ by the volume average, in a given volume of the phase space $V_6$, while $(\bar{{\bf r}},\bar{{\bf v}})$ are
the mean values of $({\bf r}_1,{\bf v}_1)$ and $({\bf r}_2,{\bf v}_2)$. The second term in Eq. (\ref{stable}) dominates in the stable clustering regime, where
$\left|\Delta {\bf v}\right| = \left|{\bf v}_1-{\bf v}_2\right| \ll \Delta v_{\rm
tid} $ and $\left|\Delta {\bf r}\right| =\left|{\bf r}_1-{\bf r}_2\right| \ll \Delta r_{\rm
tid}$, and $\Delta v_{\rm tid}$ and $\Delta r_{\rm tid}$
characterize the tidal truncation radius in the phase space. 
On the other hand, the first term dominates Eq. (\ref{stable}) for large separations in the phase space, where particles are not correlated. In other words, Eq. (\ref{stable}) is an interpolation between the stable clustering and the smooth halo regimes.  }



%
%

The annihilation signal from gravitationally bound substructure is
found by integrating (\ref{stable}) within the phase-space
stable clustering hypothesis:
\bea
\delta\Phi_{\rm sub.} &\simeq& \int
d^3{\bf x}\int d^3{\bf v}_1 d^3{\bf v}_2 \langle f({\bf x},{\bf
v}_1)\rangle \mu\xi_s(0,{\bf v_1-v_2})\nonumber\\
&=&\int d^3{\bf x}\rho_{\rm halo}({\rm x})\int d^3{\bf \Delta v}
\cdot \mu\xi_s(0,{\bf \Delta v})\;,
\eea
yielding the local boost factor of :
\bea
&&B_{\rm sub.}({\bf x}) \equiv \frac{\langle \delta \rho_{\rm
halo}({\rm x})^2 \rangle}{\langle\rho_{\rm
halo}({\rm x})\rangle^2} \simeq \frac{\mu }{\rho_{\rm
halo}({\rm x})} \int d^3{\bf \Delta v} \cdot \xi_s(0,{\bf \Delta v})
 \nonumber\\
&&\simeq \frac{8\pi^{1/2}\mu}{9\delta^3_c}\left(\rho_{\rm
halo}({\bf x})\over 200\rho_{\rm crit,0}\right)^{-1}\!\!\!\! \int^{M_{\rm
max}}_{M_{\rm min}}\!\!\!
M^{-2}d\left[M^2\sigma^3(M)\right]
\label{B_sub}
\eea
where we have used
the stable clustering framework developed in
(\ref{fxis})-(\ref{deltac}) to substitute for $\xi_s$.

The asymptotic form of CDM linear power spectrum on small scales
is (e.g. \cite{Bardeen:1985tr})
$
P(k) \propto k^{n_s-4}\ln^2(k/k_{\rm eq})\;,
$
where $n_s$
is the primordial adiabatic scalar index, and $k_{\rm eq} \simeq
0.56~ \Omega_mh^2 ~{\rm Mpc}^{-1}$ is related to the comoving scale
of the horizon at matter-radiation equality. Using this asymptotic
form, we can find an analytic approximation for the boost factor:
\bea
&&B_{\rm sub.}({\bf x}) = \mu \left(\rho_{\rm halo}({\bf
x})\over 200 \rho_{\rm crit,0}\right)^{-1}\left(\Omega_m h\over
0.27\times 0.7\right)^{\frac{3}{2}
(n_{\rm eff,8}+3)}\nonumber\\
&\times&\!\left(\sigma_8\over 0.8\right)^3\!
\!\!\left\{\!K\left[\ln^{1/2} \left(M_{\rm eq}
\over M_{\rm min}\right)\!\right]\!-\!K\left[\ln^{1/2}\!\!\left(M_{\rm eq}
\over M_{\rm max}\right)\right]\right\}\;
\label{eq:boost}
\eea
where $ M_{\rm
eq} = (2.9 \times 10^{14}~M_{\odot}) (\Omega_m h^2)^{-2}, $ is the mass
associated with the horizon scale at matter-radiation equality, $
n_{\rm eff,8} \simeq -1.68, $ is the logarithmic slope of the
linear power spectrum at $\sim 8 h^{-1} {\rm Mpc}$, and
\beq
\!\!\!K(y)\simeq {9\over 10^5}
\left(\!\frac{4y^{11}}{11}-y^{9}\!\right)\exp\left[{1\over 2}(y^2-16)(n_s-1)\right]
\eeq
This analytic approximation (\ref{eq:boost}) is within $40\%$ of the exact integral
in (\ref{B_sub}) for $10^{-15} M_{\odot} < M < 10^{10}
M_{\odot}$ and $0.95 < n_s <1.05$ (where we compare with the fitting
form of \cite{Eisenstein:1997ik} for the CDM power
spectrum). For supersymmetric dark matter models, the minimum CDM halo
mass can range from $10^{-12} M_{\odot}$-$10^{-4} M_{\odot}$,
yielding $K \simeq 10^5$ within a factor of $3$.

To calibrate the parameter $\mu$, we can compare our
results to numerical simulations. Matching the expected boost per
mass decade to that of ``Via Lactea''  mock Milky Way
simulation \cite{Diemand:2006ik}, which is around $0.1$ for sub-halo
masses $10^7 M_{\odot} < M_{sub} < 10^{10} M_{\odot}$, with their assumed cosmology,
yields
$
\mu \simeq 0.026\;\,,
$
which is comparable to the
fraction of simulated halo mass found in (resolved) gravitationally bound
sub-haloes (e.g. 5.3\% in \cite{Diemand:2006ik}).

\begin{figure}
\includegraphics[width=0.9\linewidth]{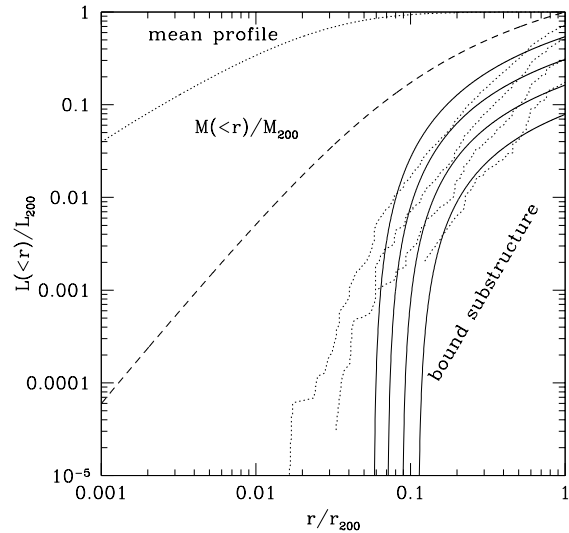}
 \caption
{The cumulative radial profile of DM annihilation: top dotted curve shows the contribution from the mean profile, while the lower dotted curves show the simulated contribution of sub-haloes more massive than $10^{5,6,7,8}\msun$ \cite{2008Natur.456...73S}. The solid curves show our analytic prediction, with the same cut-off's on $M_{\rm min}$, and a tidal cut-off at the high mass end. The dashed curve shows the cumulative mass profile for comparison.
}
\label{subhalob}
\end{figure}

Fig.(\ref{subhalob}) compares the prediction for the annihilation profile (solid curves) with numerical simulations of \cite{2008Natur.456...73S} (dotted curves), which shows reasonable agreement. To do this, we have cut off the integral (\ref{B_sub}) at high masses by requiring that the formation density of bound subhaloes must be lower than the local halo density. The crudeness of this criterion is most likely responsible for the sharp drop in the signal at small radii, relative to the numerical results. However, we note that (surviving) sub-solar mass subhaloes, that dominate the boost for realistic WIMP models, are much less affected by the tidal effects of the host halo (which was the motivation for the stable clustering approximation). Moreover, the simulated results do not include the effect of sub-sub-haloes, etc. which could be another source of difference with our prediction.

\begin{figure}
\includegraphics[width=0.9\linewidth]{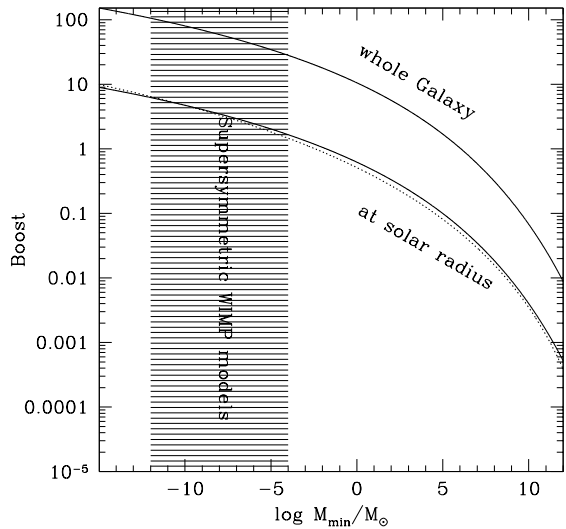}
 \caption
{Boost to the dark matter annihilation signal due to
 substructure with
the formation mass larger than $M_{\rm min}$.
 The lower solid curve is the local boost factor at solar radius
 (assuming $\rho=10^5 \rho_{\rm crit}$), while
the upper solid curve shows the mean boost estimated for
the whole Milky Way halo  (out to 16 times NFW scale radius
\cite{Diemand:2006ik}), where we have assumed $n_s=0.96$, $\sigma_8=0.82$,
$\Omega_m=0.28$, and $h=0.7$. The dotted curve is our analytic
approximation to the boost factor (\ref{eq:boost}). The dashed region
shows the theoretical expectation for the mass cut-off
\cite{Profumo:2006bv}.
}
\label{subhalot}
\end{figure}

\begin{figure}
\includegraphics[width=0.9\linewidth]{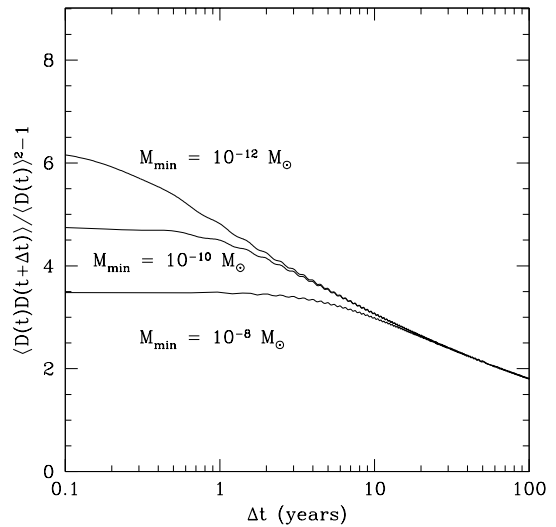}
 \caption{
Temporal correlation function of the dark matter
detection signal within the solar system
(assuming $\rho_{\rm CDM} =  10^5 \rho_{\rm crit}$) for three
different cut-offs of the CDM hierarchy.}
 \label{subhalo_directt}
\end{figure}

Fig. (\ref{subhalot}) shows that not only the CDM hierarchy can significantly
boost the total annihilation signal from the Milky Way (and dark
matter haloes in general), they can also affect the
local density variance at the solar radius ($B_{\rm sub.} \sim 2-6$), which can have
interesting implications for direct dark matter detection searches.
We can quantify the latter through the temporal correlation of dark
matter detection signal $D(t)$, which
traces the local density of the dark matter halo at the solar system.
For simplicity, we will assume that
the solar system is moving through the CDM hierarchy at $v \simeq
\sqrt{2.5} \times 250 \kms$ (assuming a singular isothermal mean phase space distribution). The
two-point correlation function
of $D(t)$ simply measures the two-point correlation of CDM density, by projecting (\ref{stable}) into
the real space, which is then modulated by the annual motion of earth around the sun. This can be done through a simple generalization of
(\ref{B_sub}) to allow for
finite separation in real space, and is shown in
Fig.(\ref{subhalo_directt}) for three different cut-offs of
the CDM hierarchy, assuming $\rho_{\rm CDM} =  10^5 \rho_{\rm crit}$
within the solar system. We thus
predict that, depending on the cut-off in the CDM hierarchy, the dark
matter detection signal could gradually
change by up to a factor of $2$ within a ten year period.
Potential measurement of this temporal correlation
could shed light on the
cut-off of the CDM hierarchy, which is
directly related to the mass of CDM particles.

\begin{figure}
\includegraphics[width=0.9\linewidth]{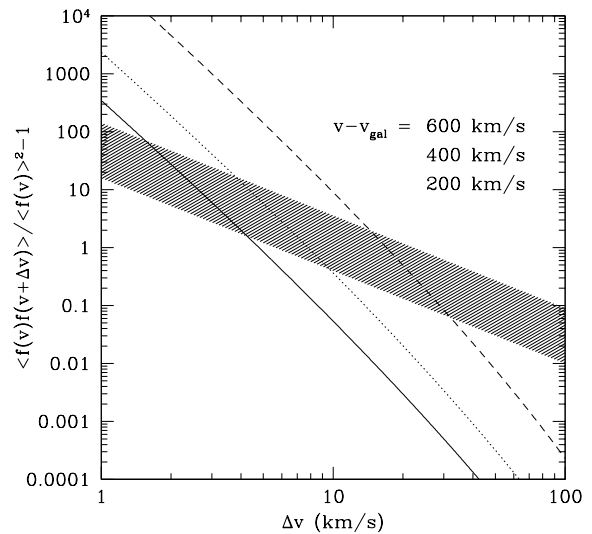}
 \caption{DM phase space correlation due to bound subhaloes
at solar radius. The three lines show velocities:
200, 400, and 600 km/s in the Galactic frame. The shaded area shows
the expected contribution from unbound substructure \cite{Afshordi:2008mx}.}
 \label{subhalo_directb}
\end{figure}

To conclude this letter, we will turn to directional dark matter searches, which can directly probe the phase space of the CDM halo. These searches are now underway, and can potentially provide the first telescopes for the rich field of {\it Dark Matter Astronomy}. We can use our formalism (Eqs. \ref{fxis}-\ref{deltac}) to predict the velocity space correlation function at the solar radius, which can be directly measured if CDM particles are detected in directional searches. This is shown in Fig. (\ref{subhalo_directb}). For simplicity, we assumed a singular isothermal sphere for the mean phase space density at the solar radius. The dimensionless correlation function is more prominent at higher velocities, and shows a correlation length of 5-20 km/s in the velocity space. For comparison, the shaded area shows the expected level of correlation due to unbound substructure, which is 10-30\% at $\Delta {\bf v} \sim 100$ km/s \cite{2009MNRAS.395..797V}, and scales as $|\Delta {\bf v}|^{-1.6}$ \cite{Afshordi:2008mx}.

The fact that $B_{\rm sub.} \sim 2-6$ implies that the local CDM density may be dominated by small subhaloes with random velocities with respect to the Galaxy. This will introduce an ${\cal O}(1)$ uncertainty in the direction of local dark matter wind (or dipole), which was believed to be the benchmark of directional dark matter searches, or the annual modulation signal. Nevertheless, as we argued above, the richness of structure introduced by the CDM hierarchy leads to novel observables/smoking guns for dark matter searches to aim for.


We thank C.-P. Ma, J. Taylor, M. Vogelsberger and S. White for
discussions. NA was partially supported by Perimeter
Institute (PI) for Theoretical Physics.  Research at PI is
supported by the Government of Canada through Industry Canada and by the
Province of Ontario through the Ministry of Research \& Innovation. RM
thanks the French ANR (OTARIE) for financial support. EB acknowledges support from NASA grant NNG06GG99G.

\vspace*{-0.5cm}


\end{document}